\documentclass[%
,aps%
 ,twocolumn%
 ,secnumarabic%
,amssymb, amsmath,  aps, floatfix, showpacs]{revtex4-1}

\begin{document}

\title{Analysis of the non-minimally coupled scalar field in the Palatini formalism by the Noether symmetry approach}
\author{Rudinei C. de Souza}
\author{Ra\'{\i}la Andr\'e}
\author{Gilberto M. Kremer}
\affiliation{Departamento de F\'{\i}sica, Universidade Federal do Paran\'a, Curitiba, Brazil}

\begin{abstract}
We analyse a scalar field non-minimally coupled to gravity in the context of a Universe described by the flat Friedmann-Robertson-Walker (F-R-W) metric. The adopted model comprises a Universe filled by the scalar field and standard matter (dark and baryonic). The corresponding field equations are obtained through the Palatini formalism. From the action of the model in the flat F-R-W space-time, a point-like Lagrangian of first order is obtained and the Noether symmetry approach is applied to restrict the forms of the \emph{a priori} undefined coupling and potential of the scalar field. We show that the massive scalar field is associated with a Noether symmetry of the model. Analytical cosmological solutions for this case are found and their respective importance for the description of the dark energy are discussed.
\end{abstract}

\maketitle

\section{Introduction}

The scalar fields play a fundamental role in several physical phenomena and are a useful tool in modern cosmology, being an indispensable part of a theoretical framework. In the model adopted in this work the scalar field acts as a source of gravity and non-minimally couples to the curvature through a generic function $F(\phi)$ and its self-interaction potential is given by a generic function $V(\phi)$.

The purpose of this paper is to analyse the dynamics of this field in the Palatini formalism \cite{ferraris82, sotiriou06, harko11} through the Noether symmetry approach \cite{capozziello07, vakili08, roshan08, desouza08, capozziello10, zhang10, basilakos11, jamil12}. The most fundamental aspect to be noted in the Palatini formalism is the independence, \emph{a priori}, between the metric tensor and affine connection, which are in fact independent geometric variables\cite{ferraris82}. The variation of the action with respect to the metric tensor yields modified Einstein's field equations and the variation with respect to the affine connection reveals the dynamic equation responsible for the new field relative to the generalized affine connection. The action is formally the same, but the new Riemann and Ricci tensors are built from the independent affine connection.

By considering a flat F-R-W metric in the action of the model, we can derive a point-like Lagrangian which furnishes the same dynamic equations that are generated from the tensor components of the modified Einstein's field equations for the respective cosmological model. Using such a point-like Lagrangian, we can restrict the forms of the undefined coupling and potential of the action by imposing that the Lagrangian presents a Noether symmetry, which formally works as a first principle for the determination of the undefined functions without \emph{ad hoc} procedures \cite{capozziello2007, desouza09, desouza10, atazadeh12, kucukakca12}. Since the dynamical system presents a Noether symmetry, there exists a constant of motion associated with and through a coordinate transformation it is possible to find a cyclic variable. Being this procedure performed, the constant of motion and the cyclic variable may lead us to the complete integration of the dynamical system \cite{capozziello00, paliathanasis11, vakili11, capozziello12, desouza13, desouza2013}. As follows, we will apply the Noether symmetry approach to determine the generic coupling and potential of the model and analytically analyse its dynamics.

This paper is organized as follows. In Section II we introduce the model in the Palatini formalism. The point-like Lagrangian for a flat F-R-W metric and the field equations are presented in Section III. In Section IV we apply the Noether symmetry approach to constrain the undefined functions. The solutions of the field equations and their cosmological meaning are presented in section V. In section VI we close the paper with the conclusions.

In this work we adopt the natural units $8\pi G=\hbar=c=1$ and the metric signature $(+, \ -, \ -, \ -)$.

\section{The model and the Palatini Formalism}

The model we consider in this work is described by the following action
\begin{align} \label{eq:1}
S=\int{d^{4}x \sqrt{-g} \left[F(\phi)\tilde{R}+\frac{1}{2} \partial_{\mu}\phi \partial^{\mu} \phi -V(\phi)+\mathcal{L}_{m}\right]},
\end{align}
where \emph{$\phi$} represents the scalar field, \emph{$V(\phi)$} is the self-interaction potential, and \emph{$F(\phi)$} denotes the coupling function between the scalar field and gravity. Here \emph{$\tilde{R}$} stands for the generalized Ricci scalar and \emph{$\mathcal{L}_{m}$} is the Lagrangian of the matter fields, comprising the ordinary and dark matter.

By varying the action with respect to the metric tensor, in accordance with the principle of least action we find the respective modified Einstein's field equations
\begin{align} \label{eq:2}
2F \left(\tilde{R}_{\mu \nu}-\frac{1}{2}g_{\mu \nu} \tilde{R} \right)=-T_{\mu \nu}\hspace{0.5mm},
\end{align}
where \emph{$T_{\mu \nu}=T_{\mu \nu}^{(\phi)}+T_{\mu \nu}^{(m)}$} is the total energy-momentum tensor. The energy momentum-tensor of the scalar field is given by
\begin{align} \label{eq:3}
T^{(\phi)}_{\mu \nu}=\partial_{\mu} \phi \partial_{\nu}\phi -\left[\frac{1}{2} \partial_{\sigma}\phi \partial^{\sigma}\phi-V(\phi)\right]\hspace{0.5mm}g_{\mu \nu},
\end{align}
and the matter energy-momentum tensor is
\begin{align} \label{eq:4}
T^{(m)}_{\mu \nu}=2\hspace{0.5mm}\frac{\partial \mathcal{L}_{m}}{\partial g_{\mu \nu}}-g_{\mu \nu} \mathcal{L}_{m}.
\end{align}

Now, by varying the action with respect to the connection through the Palatini equation
\begin{align} \label{eq:5}
\delta \tilde{R}_{\mu \nu}=\tilde{\nabla}_{\nu}\hspace{0.5mm}(\delta \tilde{\Gamma}^{k}_{\mu k})-\tilde{\nabla}_{k} \hspace{0.5mm}(\delta \tilde{\Gamma}^{k}_{\mu \nu}),
\end{align}
with $\tilde{\nabla}_{\mu}$ denoting the covariant derivative associated with $\tilde{\Gamma}^{k}_{\mu\nu}$, we obtain the equation responsible for the dynamics of \emph{$ \tilde{\Gamma}^{k}_{\mu\nu}$}, namely,
\begin{align} \label{eq:6}
\tilde{\nabla}_{k}\hspace{0.5mm}(\sqrt{-g}\hspace{0.5mm}F(\phi)\hspace{0.5mm}g^{\mu \nu})=\hspace{0.5mm}0.
\end{align}
The solution of this equation is the generalized affine connection
\begin{align} \label{eq:7}
\tilde{\Gamma}^{\sigma}_{\mu \nu}=\Gamma^{\sigma}_{\mu \nu}+\frac{1}{2}g^{k \sigma}\hspace{0.5mm}[g_{k \nu}\partial_{\mu}F+g_{\mu k}\partial_{\nu}F-g_{\mu \nu}\partial_{k}F],
\end{align}
where the $\Gamma^{\sigma}_{\mu \nu}$ are the usual Christoffel symbols.

The Ricci tensor corresponding to this generalization is given by
\begin{align} \label{eq:8}
\tilde{R}_{\mu \nu}=R_{\mu \nu}&+\frac{1}{F}\nabla_{\nu}\nabla_{\mu}F-\frac{3}{2F^{2}}\partial_{\mu}F\partial_{\nu}F\nonumber\\
&+\frac{1}{2F}g_{\mu \nu}\nabla_{\sigma}\nabla^{\sigma}F\hspace{0.5mm},
\end{align}
where \emph{$R_{\mu \nu}$} is the usual Ricci tensor and $\nabla_{\mu}$ is the covariant derivative associated with the Christoffel symbols.

If we define an effective energy-momentum tensor as
\begin{align} \nonumber\label{eq:9}
\mathcal{T}_{\mu \nu}= T_{\mu \nu}&+2 \nabla_{\nu} \nabla_{\mu}F-\frac{3}{F} \partial_{\mu}F \partial_{\nu}F-2\nabla_{\sigma}\nabla^{\sigma}F g_{\mu \nu}\\
 &+\frac{3}{2F} \partial^{\sigma}F\partial_{\sigma}F g_{\mu \nu}\hspace{0.5mm},
\end{align}
 which absorbs the non Christoffel symbols terms of the connection, we can write the field equations with the gravitational part in the standard form of the Einstein's equations
 \begin{eqnarray}\label{eq:9.1}
{R}_{\mu\nu}-\frac{1}{2}g_{\mu\nu}{R}=-\frac{\mathcal{T}_{\mu \nu}}{2F}.
\end{eqnarray}
Note that $F=1/2$ recovers the Einstein's gravitation.

From the variation of action \eqref{eq:1} with respect to the scalar field, we have the respective Klein-Gordon equation
\begin{align} \nonumber \label{eq:13}
\left(1-\frac{3F'^{2}}{F}\right)\nabla_{\mu}\nabla^{\mu}\phi &+ 3 \left(\frac{F'^{3}}{2F^{2}}-\frac{F'F''}{F}\right)\partial_{\mu}\phi \partial^{\mu}\phi\\
&-RF'+V'=0 \hspace{0.5mm}.
\end{align}

The complete dynamics of the model is then described by the field equations (\ref{eq:9.1}) and (\ref{eq:13}).
\section{Point-like Lagrangian and F-R-W field equations}

For the analysis of the cosmological aspects of the model through the Noether symmetry approach, it is a needed step to determine the point-like Lagrangian corresponding to the model for a F-R-W metric. From action (\ref{eq:1}), after eliminating the boundary terms, we obtain the point-like Lagrangian
\begin{align}\label{eq:12}
\mathcal{L}= 6Fa\dot{a}^{2}&+6F'a^2\dot{a}\dot{\phi}-\frac{a^3}{2}\left(1-\frac{3F'^{2}}{F}\right)\dot{\phi}^{2}\nonumber\\
&+a^{3}V+\rho^{0}_{m}\hspace{0.5mm},
\end{align}
where the dot represents time derivative and $\rho_m^0$ is the matter energy density at an initial instant.

From the Euler-Lagrange equation for $a$ applied to (\ref{eq:12}), we obtain the acceleration equation
\begin{align} \label{eq:14}
\frac{\ddot{a}}{a}=-\frac{\left(\rho_m+\rho_{\phi}+3p_{\phi}\right)}{12F}\hspace{0.5mm},
\end{align}
and by imposing that the energy function associated with (\ref{eq:12}) vanishes, we have de Friedmann equation, namely
\begin{align}\label{eq:14.1}
&E_{\mathcal{L}}=\frac{\partial\mathcal{L}}{\partial\dot a}\dot a+\frac{\partial\mathcal{L}}{\partial\dot\phi}\dot\phi-\mathcal{L}=0\nonumber\\ &\Longrightarrow H^{2}=\frac{\rho_m+\rho_{\phi}}{6F}\hspace{0.5mm}.
\end{align}

In equations \eqref{eq:14} and \eqref{eq:14.1} we defined the effective energy density and pressure of the scalar field as follows
\begin{align} \label{eq:10}
&\rho_{\phi}=\frac{\dot{\phi}^{2}}{2}+V-6H F'\dot{\phi}-\frac{3}{2F}F'^{2}\dot{\phi}^{2},\\
\label{eq:11}
&p_{\phi}=\frac{1}{2} \dot{\phi}^{2}-V+2 \left(F'' \dot{\phi}^{2}+2H F'\dot{\phi}+F' \ddot{\phi} \right)-\frac{3}{2F}F'^{2}\dot{\phi}^{2},\nonumber\\
\end{align}
in accordance to the energy-momentum tensors (\ref{eq:3}) and (\ref{eq:9}).

Finally, the Euler-Lagrange equation for $\phi$ applied to (\ref{eq:12}) furnishes the F-R-W Klein-Gordon equation
\begin{align}\nonumber\label{eq:126}
\left(1-\frac{3F'^{2}}{F} \right)\ddot{\phi}&+3H \left(1-\frac{3F'^{2}}{F}\right) \dot{\phi}\nonumber\\
&+3 \left(\frac{F'^{3}}{2F^{2}}-\frac{F'F''}{F}\right)\dot{\phi}^{2}-RF'+V'=0 \hspace{0.5mm}.
\end{align}

Note that equations (\ref{eq:14}), (\ref{eq:14.1}), and (\ref{eq:126}) are necessarily the same of those from the application of the F-R-W metric to the field equations (\ref{eq:9.1}) and (\ref{eq:13}), respectively. In the next section we will apply the Noether symmetry condition to the Lagrangian (\ref{eq:12}).


\section{Noether symmetry}

Now we are ready to employ the Noether symmetry approach to constrain the self-interaction potential and coupling to gravity. Let us consider the following infinitesimal generator of symmetry
\begin{align} \label{eq:15}
\textbf{X}=\alpha \frac{\partial}{\partial a}&+\beta \frac{\partial}{\partial \phi}+\left(\frac{\partial \alpha}{\partial a}\dot{a}+\frac{\partial \alpha}{\partial \phi}\dot{\phi}\right)\frac{\partial}{\partial \dot{a}}\\ \nonumber
&+\left(\frac{\partial \beta}{\partial a}\dot{a}+\frac{\partial \beta}{\partial \phi} \dot{\phi}\right)\frac{\partial}{\partial \dot{\phi}}\hspace{0.5mm},
\end{align}
where $\alpha$ and $\beta$ are functions only of $a$ and $\phi$. There will exist a Noether symmetry for the point-like Lagrangian of our model if the condition
\begin{align}\label{eq:15.1}
L_{\textbf{X}}\mathcal{L}=\textbf{X}\mathcal{L}=0
\end{align}
holds, i.e., if the Lie derivative of the Lagrangian with respect to the vector field $\textbf{X}$ vanishes \cite{rittis90, capozziello93, capozziello2008, desouza11}.
By applying the symmetry condition (\ref{eq:15.1}) to (\ref{eq:12}), with respect to the vector field (\ref{eq:15}), we obtain a system of coupled partial differential equations as shown bellow
\begin{align}\label{eq:16}
&3 \alpha V+a \beta V'=0 \hspace{0.5mm},\\
\label{eq:17}
&\alpha+2a\frac{\partial \alpha}{\partial a}+a \left(\beta+a \frac{\partial \beta}{\partial a}\right)\frac{F'}{F}=0 \hspace{0.5mm},\\
 \nonumber\label{eq:18}
&\beta a F''+\left(2 \alpha+a \frac{\partial \alpha}{\partial a}+a \frac{\partial \beta}{\partial \phi}\right)F'+2F\frac{\partial \alpha}{\partial \phi}\\ &\ \ \ \ \ \ \ \ \ \ \ \ \ \ \ \ \ \ \ \ \ \ \ \ \ \ -\frac{a^{2}}{6} \left(1-\frac{3F'^{2}}{F}\right)\frac{\partial \beta}{\partial a}=0 \hspace{0.5mm},\\
 \nonumber\label{eq:19}
&3 \alpha-12 \frac{\partial \alpha}{\partial \phi}F'+2a \left(1-\frac{3F'^{2}}{F}\right)\frac{\partial \beta}{\partial \phi}\\
&\ \ \ \ -3 \left(3 \alpha-a \beta \frac{F'}{F}+2 a \beta \frac{F''}{F'}\right)\frac{F'^{2}}{F}=0 \hspace{0.5mm}.
\end{align}

 To solve the above system we consider $\alpha$ and $\beta$ as separable functions of $a$ and $\phi$, i.e., $\alpha=\alpha_1(a)\alpha_2(\phi)$ and $\beta=\beta_1(a)\beta_2(\phi)$. The respective solutions are in Table 1 which contains all the sets of functions \emph{$\alpha$}, \emph{$\beta$}, \emph{$V(\phi)$}, and \emph{$F(\phi)$}, where the quantities $m, \alpha_{0}, V_0$ and $F_0$ are constants. If we now employ the potentials and couplings of Table 1, the model will present a Noether symmetry.
\begin{table*}
   \begin{tabular}{ | l | l | l | l | l | l | p{10cm} |}
    \hline
    Range of validity & $\alpha$ & $\beta$ & $V$ & $F$ & $F_{0}$ \\ \hline
    $m$, $n \neq 0$; $m \neq -\frac{1}{2}$  & $\alpha_{0}a^{m} \phi^{n}$ & $-\left(3 \alpha_{0}/2l \right)a^{m-1} \phi^{n+1}$ & $V_{0}F^{\frac{3m}{2m+1}}$ & $F_{0}\phi^{2}$ & $\frac{2m-n-2mn-2m^{2}}{24(m-n-1)}$ \\ \hline
    $m=1$, $n=0$ & $\alpha_{0}a$ & $-(3 \alpha_{0}/2)\phi$ & $V_{0}\phi^{2}$ & $F_{0}\phi^{2}$ & $F_{0}$ \\
    \hline
    \end{tabular}
\begin{center}\textbf{Table 1.} Set of solutions.\end{center}
\end{table*}

We are interested in working with the solution for $m=1$ and $n=0$ since the respective potential describes a massive scalar field, which is physically interesting. The conserved quantity associated with the Noether symmetry corresponding to this solution reads
\begin{align}\label{eq:20}
\Sigma_{0}=-6F_{0}\phi a^{2} \left(\phi \dot{a}+a \dot{\phi} \right)+\frac{3}{2}a^{3}\phi \dot{\phi}\hspace{0.5mm}.
\end{align}

As follows, we will search for analytical solutions for the case of the massive scalar field.
\section{Solutions of the field equations}

To find the solutions of the field equations we need to rewrite the point-like Lagrangian \eqref{eq:12} in other system of coordinates which makes integration easier. Thus, by knowing that there is a Noether symmetry related to \emph{$F$} and \emph{$V$}, there must exist a coordinate transformation in the space of configuration in which one of these coordinates is cyclic. Such a transformation obeys the following system of differential equations
\begin{align} \label{eq:21}
&\alpha \frac{\partial u}{\partial a}+\beta \frac{\partial u}{\partial \phi}=0,\\
&\alpha \frac{\partial z}{\partial a}+\beta \frac{\partial z}{\partial \phi}=1,
\end{align}
where $u$ and $z$ are the new coordinates linked to the old ones, $a$ and $\phi$. In this transformation $z$ is the cyclic coordinate.
A convenient particular solution of this system for the case $m=1, n=0$ of Table 1 is
\begin{align} \label{eq:21.1}
u=a^{3}\phi^{2}, \qquad z=\ln a.
\end{align}

The point-like Lagrangian (\ref{eq:12}), with $F$ and $V$ respective to the case $m=1, n=0$, rewritten in the new variables given by (\ref{eq:21.1}) takes the form
\begin{align} \label{lag}
\mathcal{L}=k_{1}\dot{u}\dot{z}-k_{2}u \dot{z}^{2}+k_{3}\frac{\dot{u}^{2}}{u}+V_{0}u+\rho^{0}_{m},
\end{align}
with
\begin{align} \label{coeff}
&k_{1}=\frac{3}{4}\left(1-4F_{0}\right), \quad
k_{2}=\frac{3}{8}\left(3-4F_{0}\right),\nonumber\\
&k_{3}=-\frac{1}{8}\left(1-12F_{0}\right).
\end{align}

Now, from the Euler-Lagrange equations associated with the Lagrangian \eqref{lag}, we obtain the field equations in the new variables, namely,
\begin{align}\label{eq:176}
&2k_{2}u\dot{z}-k_{1}\dot{u}=\Sigma_{0},\\
\label{eq:177}
&2 k_{3}\frac{\ddot{u}}{u}+k_{1}\ddot{z}+k_{2}\dot{z}^{2}-k_{3}\frac{\dot{u}^{2}}{u^{2}}-V_{0}=0,
\end{align}
where \emph{$\Sigma_{0}$} is nothing else than the constant of motion \eqref{eq:20} rewritten in the new variables.

The energy function associated with the Lagrangian \eqref{lag} provides another equation
\begin{align}\label{eq:178}
k_{3}\frac{\dot{u}^{2}}{u^{2}}+k_{1}\frac{\dot{u}}{u}\dot{z}-\left(V_{0}+k_{2}\dot{z}^{2}\right)-\frac{\rho^{0}_{m}}{u}=0,
\end{align}
which is equivalent to the Friedmann equation in the old variables.

The system (\ref{eq:176})-(\ref{eq:178}) comprehends three differential equations for two dynamical variables, $u$ and $z$. Then one can solve it completely from any pair of equations. For this task we take equations (\ref{eq:176}) and (\ref{eq:178}), which are simpler to solve once they involve only first order derivatives. Proceeding in this way, we obtain two solutions for the system. The first one, expressed in the original variables, reads
\begin{align}
&a(t)=a_0\left[\frac{\tanh{\frac{1}{2}\left(u_{1}t+u_2\right)+u_{0}-\sqrt{1+u_{0}^{2}}}}{\tanh{\frac{1}{2} \left(u_1 t+u_2\right)+u_{0}+\sqrt{1+u_{0}^{2}}}}\right]^{p} \nonumber\\
& \ \ \ \ \ \ \ \ \ \ \ \ \ \ \ \ \ \ \times\left[u_{0} \sinh{ \left(u_{1}t+u_2\right)}-1\right]^{\frac{k_{1}}{2k_{2}}},\label{eq:22}\\
&\phi(t)=\phi_{0}\left[\frac{\tanh{\frac{1}{2}\left(u_{1}t+u_2\right)+u_{0}-\sqrt{1+u_{0}^{2}}}}{\tanh{\frac{1}{2} \left(u_1 t+u_2\right)+u_{0}+\sqrt{1+u_{0}^{2}}}}\right]^{-\frac{3p}{2}}\nonumber\\
& \ \ \ \ \ \ \ \ \ \ \ \ \ \ \ \ \ \ \times\left[u_{0} \sinh{\left(u_{1}t+u_2\right)}-1\right]^{\frac{2k_{2}-3k_{1}}{4k_{2}}},\label{eq:23}
\end{align}
where $u_2$ is a constant of integration and
\begin{align}
&a_0=z_0\left(\frac{\rho_m^0}{2V_0}\right)^\frac{k_1}{2k_2}, \qquad u_{1}=2\sqrt{\frac{k_{2}V_{0}}{k_{1}^{2}+4k_{2}k_{3}}},\nonumber\\
&u_{0}=\sqrt{\frac{\Sigma^{2}_{0}V_{0}}{k_{2}\left(\rho^{0}_{m}\right)^{2}}-1}, \qquad p=\frac{\Sigma_{0}V_{0}}{\rho^{0}_{m}k_{2}u_1}\frac{1}{\sqrt{1+u_{0}^{2}}},\nonumber\\
 &\phi_{0}=z_{0}^{-3/2} \left(\frac{\rho^{0}_{m}}{2V_{0}} \right)^{\frac{2k_{2}-3k_{1}}{4k_{2}}},
\end{align}
with $z_0$ being another constant of integration. This solution is valid for
\begin{equation}
\Sigma^{2}_{0}>\frac{k_{2}\left(\rho^{0}_{m}\right)^{2}}{V_{0}}.
\end{equation}

The second one reads
\begin{align}\label{eq:24}
&a(t)=a_{0}\hspace{0.5mm} e^{-a_{1}t} \left(u_{3}e^{u_{1}t}-1 \right)^{q+\frac{k_{1}}{2k_{2}}},\\
&\phi(t)=\phi_{0}\hspace{0.5mm} e^{\frac{3a_{1}}{2}t} \left(u_{3} e^{u_{1}t}-1 \right)^{\frac{2\left(1-3q \right)k_{2}-3 k_{1}}{4 k_{2}}},
\end{align}
where
\begin{align}
&a_1=\frac{\Sigma_0V_0}{\rho_m^0k_2}, \qquad u_3=\frac{V_0e^{u_2}}{\rho_m^0}, \qquad q=\frac{\Sigma_0V_0}{\rho_m^0k_2u_1}.
\end{align}
Such a solution is valid for
\begin{equation}
\Sigma^{2}_{0}=\frac{k_{2}\left(\rho^{0}_{m}\right)^{2}}{V_{0}}.\label{condition}
\end{equation}

Let us analyse the cosmological meaning of the found solutions by means of their asymptotic behaviors. If we set $u_2=0$ in solution (\ref{eq:22}) and expand the hyperbolic functions up first order in $t$, by considering $u_1t\ll u_0$ for small $t$, i.e., that $u_0$ is sufficiently larger than unity, one has the respective asymptotic form
\begin{align}
&a(t)=a_0\left(u_{0}u_{1}\right)^{\frac{k_{1}}{2k_{2}}}\left(\frac{u_{0}-\sqrt{1+u_{0}^{2}}}{u_{0}+\sqrt{1+u_{0}^{2}}}\right)^{p}
\left(t-\frac{1}{u_{0}u_{1}}\right)^{\frac{k_{1}}{2k_{2}}},
\end{align}
which is valid for small $t$. In this asymptotic limit we require that the usual matter era holds, which occurs if the exponent \emph{${k_{1}}/{2k_{2}}$} presents the value \emph{${2}/{3}$}. On the other hand, from (\ref{coeff}) we see that such a requirement implies that \emph{$F_{0}=-3/4$}. Thus the usual matter era is not possible in this solution since gravity must be attractive. By requiring that \emph{${k_{1}}/{2k_{2}}>0$} -- a non-collapsing Universe -- at the same time that $F_0>0$, we have that the exponent must be limited in the range \emph{$0<{k_{1}}/{2k_{2}}<{1}/{3}$} and $0<F_0<1/4$. Hence the model can produce a decelerated expansion in the past although not a usual matter dominated era, with the effective fluid composed of scalar and matter fields driving such a deceleration.

Under the same considerations, the asymptotic behavior of solution (\ref{eq:22}) for large $t$ is
\begin{align}
&a(t)=a_0\left(\frac{u_{0}}{2}\right)^{\frac{k_{1}}{2k_{2}}}
\left(\frac{1+u_{0}-\sqrt{1+u_{0}^{2}}}{1+u_{0}+\sqrt{1+u_{0}^{2}}}\right)^{p}e^{\frac{k_{1}u_{1}}{2k_{2}}t},
\end{align}
which stands for a de Sitter-like expansion in the future. Therefore the referred solution describes a Universe that is decelerated in the past and passes through a transition from a decelerated to an accelerated phase in the present, evolving to a de Sitter Universe in the distant future.

Proceeding as above, the asymptotic form of solution (\ref{eq:24}) for small $t$ reads
\begin{align}
&a(t)=a_{0}\left(u_1u_{3}\right)^{q+\frac{k_{1}}{2k_{2}}}\left(t-\frac{1-u_3}{u_1u_{3}}\right)^{q+\frac{k_{1}}{2k_{2}}}.
\end{align}
The usual matter era exponent, $q+{k_{1}}/{2k_{2}}=2/3$, holds in this asymptotic form in the limit $F_0\rightarrow3/4$. Note that if we set $F_0=3/4$, one has an indeterminate form in the exponent and the constant $u_1$ vanishes, freezing out the time evolution of the scale factor. Thus one can approximately recover the usual matter era in the past with $F_0$ near $3/4$, which from (\ref{condition}) implies that the constant of motion must have a small value. It is important to point out that all the positive values for $F_0$ -- except $3/4$ -- produce a non-collapsing Universe, being all of them acceptable, but only values around $F_0=3/4$ can produce a \emph{quasi} usual matter dominated era.

For large $t$, solution (\ref{eq:24}) presents the following asymptotic behavior
\begin{align}
&a(t)=a_{0}u_3^{q+\frac{k_{1}}{2k_{2}}}e^{\left[u_{1}\left(q+\frac{k_{1}}{2k_{2}}\right)-a_{1}\right]t}.
\end{align}
Keeping all the fixed considerations for small $t$, the above limit for large $t$ also generates a de Sitter-like expansion. Hence this solution also presents the property of transition from a decelerated to an accelerated phase, as well as solution (\ref{eq:22}).

Having in mind the relatively recent astronomical observations on the decelerated-accelerated expansion of the Universe \cite{riess98, perlmutter99, leibundgut01}, the general behavior of the solutions are in according with, however the first solution cannot recover the usual matter era in the past, which is approximately achieved by the second one. Thus the analysed model may more successfully reproduce the cosmological dynamics through solution (\ref{eq:24}). From these results, the massive scalar field of the model may simulate the dark energy.

\section{Conclusions}

In this work we analysed a model with a scalar field non-minimally coupled to gravity through the Palatini formalism and Noether symmetry approach and investigated its cosmological relevance for the late Universe. A technical advantage of this approach is that once the Noether symmetry exists for the point-like Lagrangian of the model, the associated constant of motion and cyclic variable can help with the integration of the system. We showed that the massive scalar field is associated with a Noether symmetry of the model and the corresponding dynamical system can be completely integrated. The solutions of the system can describe the transition from a decelerated to an accelerated expansion of the Universe. However, the first found solution cannot account for the usual matter dominated era in the past, which is approximately accounted for the second one, being more according to the astronomical observations. The both solutions forecast a de Sitter Universe in the distant future, i.e., an eternal accelerated expansion. In view of these results, the massive scalar field may play the role of dark energy in this model.


\section*{Acknowledgments}
The authors acknowledge the financial support of CNPq and Capes (Brazil).

\end{document}